\documentclass[aps,prb,twocolumn,showpacs]{revtex4}
\usepackage{tabularx,graphicx}
\newcommand{\nin}{\noindent}
\newcommand{\dlE}{\Delta \! E}

\begin{document}
\title{Electronic lifetimes in ballistic quantum dots
electrostatically\\ coupled to metallic environments}
\author{Francisco Guinea$^1$, Rodolfo A. Jalabert$^2$, and Fernando Sols$^3$}

\affiliation{$^1$ Instituto de Ciencias Materiales, CSIC,
Campus de Cantoblanco, E-28043 Madrid, Spain \\
$^2$ Institut de Physique et Chimie des Mat\'{e}riaux de Strasbourg,
             UMR 7504 (CNRS-ULP) 23 rue du Loess,
             BP 43 67034 Strasbourg Cedex 2, France \\
$^3$ Departamento de F\'{\i}sica Te\'orica de la Materia
Condensada e Instituto ``Nicol\'as Cabrera''
\\ Universidad Aut\'onoma de Madrid, Cantoblanco, E-28043 Madrid, Spain}


\begin{abstract}
We calculate the lifetime of low-energy electronic excitations in a
two-dimensional quantum dot near a metallic gate. We find
different behaviors depending on the relative values of the dot
size, the dot-gate distance and the Thomas-Fermi screening length
within the dot. The standard Fermi liquid behavior is obtained when
the dot-gate distance is much shorter than the dot size or when it
is so large that intrinsic effects dominate. Departures from the
Fermi liquid behavior are found in the unscreened dipole case of small
dots far away from the gate, for which a Caldeira-Leggett model is
applicable. At intermediate distances, a marginal Fermi liquid is
obtained if there is sufficient screening within the dot. In these
last two non-trivial cases, the level width decays as a power law
with the dot-gate distance.

\end{abstract}
 \maketitle
\section{Introduction.}
The understanding of lifetimes of electronic states in finite
low-dimensional devices presents subtleties not found in extended
systems\cite{SIA94,ISS97,AGKL97,S97}. The lifetime of
quasiparticle excitations (with energy $E$) in bulk three- and
two-dimensional clean electron systems scales as $( E - E_F )^2$,
for $E$ sufficiently close to the Fermi energy, $E_F$
\cite{N89,AFS82}. Such a behavior results from the combined
effects of the screened electron-electron interaction and the
continuous spectral density provided by the internal electronic
environment (Fermi sea). The above energy dependence can be
readily obtained from a Fermi Golden Rule (FGR) approach using the
matrix elements of the effective interaction expressed in terms of
the polarizability of the metal\cite{N89}. This response function
can also be viewed as resulting from an external environment.

The corrections induced by disorder and finite size can be
systematically computed using an expansion in the inverse
conductance, $g^{-1} = \Delta_0 / E_c$, where $\Delta_0$ is the
typical level spacing, and $E_c = \hbar D/ L^2$ is the Thouless
energy ($D$ is the diffusion coefficient, and $L$ the size of the
system)\cite{A00,ABG02}. The application of a FGR approach for
intrinsic quasiparticle decay is valid for sizes (or quasiparticle
energies) large enough that the system can be described as an
environment with a continuous spectrum\cite{AGKL97,S97}. On the
other hand, in very small dots, it is the presence of nearby
metallic gates that ensures the existence of decay channels at
arbitrarily low energies. This effect becomes of crucial
importance in determining the width of the low lying electronic
excitations \cite{Ketal97,KAT01}.

\begin{figure}
\includegraphics[angle=0,width=\columnwidth]{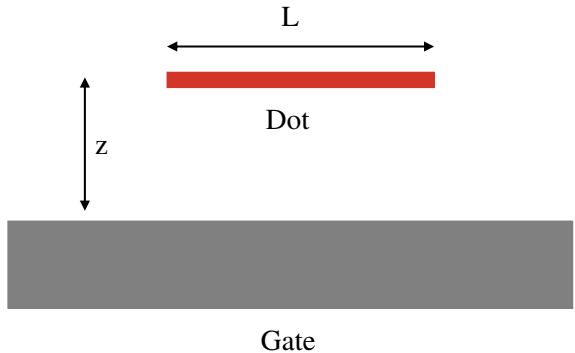}
\caption{Sketch of system studied in the text.} \label{sketch}
\end{figure}

The purpose of the present work is to investigate the
quasiparticle lifetimes in quantum dots of various sizes in the
vicinity of an extended metallic gate. In this way, we can study
the crossover between the regimes in which the external
environment does not introduce qualitative changes with respect to
the case of the isolated dot, and that in which significant
departures from Fermi liquid behavior are obtained. We consider
the simple geometry sketched in Fig. \ref{sketch}, with a
two-dimensional dot of lateral size $L$, at a distance $z$ from
the gate. Another important scale is the the Thomas-Fermi
screening length for the dot, $\lambda_{\rm TF} = \pi \hbar^2
\epsilon_0  / m^* e^2 $ ($m^*$ is the electronic effective mass
and $\epsilon_0$ the dielectric constant). The metallic gate will
be described as an Ohmic environment, that is, as one with a
low-energy spectral density proportional to the frequency
\cite{W93}. The different behaviors mentioned above are obtained
by varying the relative values of the length scales $L$, $z$, and
$\lambda_{\rm TF}$.

The widely used Caldeira-Leggett model, understood as that with
Ohmic spectral density and coupling linear in the particle
coordinate\cite{CL81}, corresponds to the limiting case $z / L ,
\lambda_{\rm TF} / L \gg 1$. This is the regime which exhibits the
largest departure from the Fermi liquid behavior, a result which
is consistent with the decay of persistent currents predicted for
mesoscopic rings where the electronic dipoles couple to an Ohmic
oscillator bath \cite{cedr01}. Clarifying the applicability and
physical consequences of different environments is one of the
guiding lines of this work.

The question of electron decay is related to the loss of phase
coherence suffered by electrons in a dot, a problem which has been
the object of experimental attention
\cite{yaco95,clar95,buks98,folk01}. Here we propose a framework
within which one can study a variety of geometrical setups where
extrinsic mechanisms compete with the intrinsic ones to cause
electron wave decoherence.

The next section introduces the model considered in this work. Then,
we present calculations for the case $z / L , \lambda_{\rm TF} / L \gg 1$.
Section \ref{sec:shortrange} discusses the regime $z / L \ll 1$, while in
Sec.~\ref{sec:screeneddip} we analyze the situation when $z / L \gg 1$ and
$\lambda_{\rm TF} / L \ll 1$. The main conclusions, as well as the
experimental relevance of our work, are discussed in
Sec.~\ref{sec:conclusion}.

\section{The model.}
\label{sec:model}

As described in the introduction, our system is given by a dot
coupled to a metallic gate (see Fig. \ref{sketch}). Such an
electrostatic interaction is governed by the Hamiltonian

\begin{equation}
{\cal H}_{\rm int} = \int_A \hat{V}_{z} ( {\bf r} ) \hat{\rho} ( {\bf
r} ) d {\bf r} \ ,
\label{hamil}
\end{equation}

\nin where $A = L^2$ is the area of the dot, ${\bf r}$ is a two-dimensional
vector in the plane of the dot, $\hat{\rho} ( {\bf r} )$ is the operator
describing the local electronic density fluctuations of the dot,
and $\hat{V}_{z}({\bf r})$ represents the potential induced on the dot by
the gate (with $z$ the separation dot-gate).

The electronic states of the dot in the absence of the gate will
be assumed to be well described by an independent electron
analysis. The consistency of this approximation in the various
regimes will be discussed in the sequel. Using the FGR
the probability per unit time of a transition between the
electronic states $n$ and $m$ (transition rate) is:
\begin{equation}
\Gamma_{mn} = \int_A d {\bf r} \int_A d {\bf r}' \langle m |
\hat{\rho} ( {\bf r} ) | n \rangle \langle n | \hat{\rho} ( {\bf
r}' ) | m \rangle S ( {\bf r}, {\bf r}',z;E_n - E_m ),
\label{lifetimes}
\end{equation}
where the structure factor
\begin{equation}
S ( {\bf r} , {\bf r}',z;\dlE) = \int \frac{d t}{\hbar} \,
e^{i \dlE t/\hbar} \left\langle \hat{V}_{z} ( {\bf r}' , t ) \hat{V}_{z}
({\bf r} , 0 ) \right\rangle \label{environment}
\end{equation}

\begin{figure}
\resizebox{6cm}{!}{\includegraphics[width=7cm]{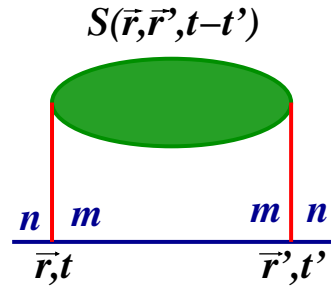}}
\caption{Diagram representing the calculation in Eq.
(\protect{\ref{lifetimes}}). The filled bubble describes the
time-dependent response of the environment (improper
polarization). Labels $m$ and $n$ stand for electron states within
the dot.} \label{diagram}
\end{figure}

\nin describes the fluctuations of the metallic environment as
experienced in the quantum dot. The averages are taken with
respect to the degrees of freedom of the environment. The second
order processes involved in the calculation of the transition
rates of Eq. (\ref{lifetimes}) are represented by the diagram
shown in Fig. \ref{diagram}.

Within the one-particle picture that we adopted for the dot
$\langle m | \hat{\rho} ({\bf r}) | n \rangle =
\Psi_m^*({\bf r})\Psi_n(\bf r)$, where $\Psi_n$ is the wave function of
the effectively independent electron, and Eq.~(\ref{lifetimes}) can be
written as

\begin{equation} \label{general-gamma-mn}
\Gamma_{mn}= \int \frac{d^2q}{(2\pi)^2}\left|\langle m |e^{i{\bf
q} \cdot {\bf r}}|n \rangle \right|^2 S({\bf q},z;\dlE)   \,,
\end{equation}

\nin where $S({\bf q},z,\dlE)$ is the Fourier transform of the
structure factor.

Notice that now a quasi-particle does not decay into a
two-particle--one-hole configuration (like in the case of
intrinsic decay), but into another one-particle state. The use of
FGR is then justified by the large density of final states
provided by the environment (metallic gate).

The properties of the environment determine the precise form of
the bubble in Fig. \ref{diagram}. However, we can infer some
general features fulfilled by the physically relevant environments
we are interested in:

\nin (i) An Ohmic environment implies that, at low energies,

\begin{equation} \label{linear-E}
S({\bf r},{\bf r}',z;\dlE) \simeq | \dlE |\ \tilde{S}({\bf r},{\bf r}',z) \ ;
\label{structure}
\end{equation}

\nin (ii) For small $| {\bf r} |$ and $| {\bf r}' |$ we will
always have, effectively,

\begin{equation}
\tilde{S}({\bf r},{\bf r}',z) \simeq {\bf r}
\stackrel{\leftrightarrow}{{S}_0}\!(z) \ {\bf r}' \ ; \label{dipolar}
\end{equation}

\nin (iii) $\tilde{S}({\bf r},{\bf r}',z) \rightarrow 0$ for $| {\bf
r} - {\bf r}' |$ sufficiently large with respect to the
characteristic length scales of the environment and the dot-gate
coupling. These general properties still allow for different
regimes, which will be thoroughly discussed in the following
chapters.

The lifetime of a state $n$ is obtained by considering all the
possible transitions to lower states above the Fermi energy:
\begin{equation}
\Gamma_n = \sum_{E_F < E_m < E_n} \Gamma_{mn} \ .
\end{equation}
We are interested in describing the energy dependence of the level
width. Therefore, we will average over nearby eigenstates:

\begin{equation}
\Gamma  ( E ) = \frac{1}{\nu(E)}\sum_n \Gamma_n \delta_\epsilon (
E - E_n ) \ , \label{lifetime}
\end{equation}

\nin where $\delta_\epsilon$ represents a smoothed $\delta$
function of width $\epsilon$ (which we take as an energy scale of
the order of a few level spacings) and $\nu(E)\equiv \sum_n
\delta_{\epsilon}(E-E_n)$ is the smoothed density of states within
the dot.

\section{Dipolar approximation.}
\label{sec:dipole}

For dots far away from the gate ($z \gg L$) and $L$ so small that
internal screening effects can be neglected ($L \ll \lambda_{\rm
TF}$), the electric field penetrates the dot and, in the
low-frequency limit, we can use the dipolar expansion represented
in Eq. (\ref{dipolar}). For simplicity, we will take the tensor
$\stackrel{\leftrightarrow}{{S}_0}$ as diagonal and isotropic,
$\stackrel{\leftrightarrow}{{S}_0}\simeq S_0(z)$. Relaxing this
assumption would not lead to qualitative changes in the results.
An estimation of ${S}_0(z)$ in a simple situation is given in
Appendix A.

From equations (\ref{lifetimes}), (\ref{linear-E}), and
(\ref{dipolar}) one can write the transition rate as

\begin{widetext}
\begin{equation}
\Gamma_{mn} = {S}_0(z)\ (E_n - E_m ) \int d {\bf r} \, d {\bf
r}'\, \Psi_n^* ( {\bf r} ) \Psi_m ( {\bf r}' ) \,{\bf r} \cdot
{\bf r}'\, \Psi_m( {\bf r} ) \Psi_n^* ( {\bf r}' ) =
{S}_0(z)\ (E_n - E_m )|\langle m |{\bf r})| n \rangle|^{2}
\ .
\label{matrix_1}
\end{equation}
\end{widetext}

The transition rate is then governed by the squared dipole matrix element.
Local averages (on a scale of $\epsilon$) of matrix elements for an
arbitrary operator $A$,

\begin{equation} \label{avmatel}
C_{A}(E,\dlE) \equiv \sum_{mn} |\langle m|A|n \rangle|^2
\delta_\epsilon (E-E_n) \delta_\epsilon (\dlE-E_n+E_m)
 \ ,
\end{equation}

\nin have been thoroughly studied \cite{FP86,W87,EFMW92,MBM95}. They
are interesting quantum mechanical quantities governing not only transition
rates, but other experimentally revelant phenomena, like the energy
absorption rate of small metallic clusters \cite{AW93,SM01} and quantum
transport in ballistic systems \cite{Kbook}. Similarly to the density of
states \cite{cit-Gutz}, $C_{A}(E,\dlE)$ can be written as a leading smooth
term plus oscillating corrections. The later are given by a periodic orbit
expansion, and therefore depend on the nature of the underlying classical
dynamics (i.e. chaotic vs. integrable). The smooth term is given by the
Fourier transform of the classical autocorrelation function of the
observable (expressed in terms of classical trajectories) and presents
some scaling behavior in the case of billiard systems \cite{MBM95}. For
the dipole matrix elements ($A={\bf r}$) the hard wall confining potential
translates into a dependence

\begin{equation} \label{correlation}
C_{\bf r}(E,\dlE)
\simeq \frac{L^2}{\sqrt{\nu ( E )}}\frac{E^{3/2}}{\dlE^4}
 \ ,
\end{equation}

\nin for both, integrable and non-integrable systems, provided
$\dlE > \hbar v_F/L$. The $\dlE^{-4}$ dependence of $C_{\bf r}(E,\dlE)$
arises from the discontinuity of the velocity of the particle as it bounces
against the hard wall. The power-law of Eq.~(\ref{correlation}) is obtained
from the contribution of a single bounce off the boundary. At sufficiently
low frequencies, the contributions from many such bounces have to be added.
This leads to interference effects and to an effective lower cutoff of the
$\dlE^{-4}$ law at the synchrotron energy  $\hbar v_F / L$.

Like in the one-dimensional case (where the semiclassical dipole matrix
element can be evaluated on general grounds\cite{IJR01}), the $\dlE^{-4}$
law is preserved when we consider a soft potential, provided that
$\dlE \ll (\hbar / a) \sqrt{E/m^*}$, where $a$ is the length scale defining
the rise of the confining potential. That is when the small energy
differences that we are interested in correspond to times much larger than
the collision time, and thus the detailed profile of the confining
walls becomes irrelevant. For large values of $\dlE$, the soft
character of the walls comes into play, and the correlation
function defined in Eq. (\ref{avmatel}) decays faster than
$\dlE^{-4}$. In the following calculations we will be using
Eq.~(\ref{correlation}) for $\dlE$ in the range $\hbar v_F/L < \dlE
<(\hbar/a)\sqrt{E/m^*}$.

In simple two-dimensional geometries, like a circular disc \cite{AW93}
or an infinite rectangular stripe \cite{SM01}, the dipole matrix
element can be calculated explicitly for the unscreened case
and screened (Thomas-Fermi) cases. In the calculation that follows
we will only need the scaling behavior of Eq.~(\ref{correlation}) and
we will not be restricted to a particular geometry.

If $\Delta_0 \approx \hbar^2 / m L^2 $ is the mean level spacing
within the dot, the estimation of the energy dependence of the
dipole matrix elements allows us to write the level width as
\begin{widetext}
\begin{equation}
\Gamma ( E )  \simeq \frac{S_0(z)}{\nu(E)} \int_{\Delta_0}^{E -
E_F} d (\dlE)\,\dlE \, C_{\bf r}(E,\dlE) \simeq c \left[1 - \left(
\frac{\hbar v_F/L}{E - E_F} \right)^2\right]
\label{gamma_dipole}
\end{equation}
\end{widetext}

\begin{figure}[b]
\resizebox{7cm}{!}{\includegraphics[width=12cm,height=10cm]{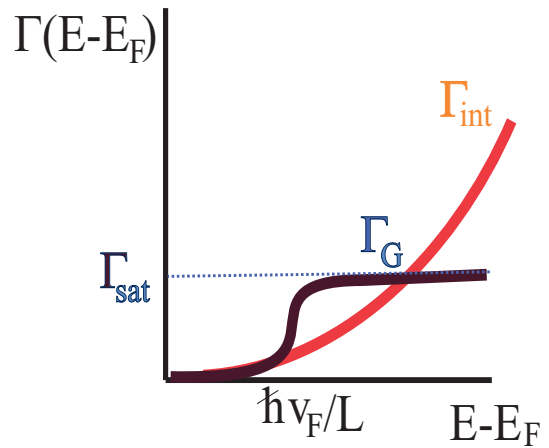}}
\caption{Schematic behavior of the decay rate, $\Gamma(E)$, of
state of energy $E$, as a function of $E$, for the unscreened
dipole case discussed in Sec. III.} \label{tau}
\end{figure}

\nin for $E-E_F > \hbar v_F/L$, and $\Gamma(E)\simeq 0$ for
$E-E_F \ll \hbar v_F/L$. The constant $c$ is discussed in the sequel. The
energy dependence of the level width, depicted in Fig. \ref{tau}, yields
a plateau-like behavior for $ \hbar v_F / L \ll E-E_F \ll E_F$.
For $E - E_F$ comparable to $E_F$, the possible smoothing of the
potential and the opening of additional relaxation mechanisms lead
to a vanishing life-time, and therefore $\Gamma (E)$ increases with $E$
for $E-E_F > E_F$ (not shown in Fig. \ref{tau}).

The quasiparticle decay rate for a standard Fermi liquid
follows the law $\Gamma_{\rm FL}(E) \propto ( E - E_F )^2$. Thus,
the energy dependence obtained in Eq. (\ref{gamma_dipole})
(a constant $\Gamma(E)$ for $E-E_F \gg \hbar v_F/L$)
reveals a {\it non--Fermi-liquid behavior}. We have thus
identified an electronic system where dissipation effects are well
described by the standard (linear in the particle coordinate)
Caldeira-Leggett model, the oscillator bath being formed by the
quasiparticle field of the gate. Anomalous behaviors in systems
described by the the Caldeira-Leggett model have been discussed in
the literature\cite{GZ98,G02}. Such particle-bath couplings were
known to be realized in contexts where the particle coordinate
(e.g. the flux through a superconducting ring) does not experience
any quantum statistical constraint. Here we have shown that, under
specific circumstances, such anomalous behavior can also be
displayed in systems where the Pauli exclusion principle plays an
essential role.

The constant $c$ in Eq.~(\ref{gamma_dipole}) depends on the distance
gate-dot and on the properties of the gate (through ${S}_0(z)$), as well
as on the details of the dot (through $C(E,\dlE)$ and $\nu(E)$). Using
the results of the Appendix for a two-dimensional gate, we have
$c=(L\hbar^2 k_F/m)z^{-2}(k_F' l')^{-1}$,
where the primes refer to parameters of the gate. For a
semi-infinite three-dimensional gate, the factor $z^2 k_F'l'$ in
the denominator of ${S}_0(z)$ is replaced by $z^3k_F'^{2}l'$, while for
a one-dimensional gate it is replaced by $zl'$. Thus, we find that the
level width decays with the dot-gate distance as power-law, with an
exponent given by the dimensionality of the gate ($d=1,2,3$)

\section{Short range regime.}
\label{sec:shortrange}

For dots sufficiently close to the gate ($z \ll L$), we expect the
quasi-perfect screening provided by the gate to yield a structure
factor of Eq. (\ref{structure}) of limited range. Let us define
${\bf r}'' = {\bf r} - {\bf {r}'}$ and assume that
\begin{equation} \label{bar-S}
\bar{S} \equiv \int d {\bf {r}}''  {S} ( {\bf {r}}'' )
\end{equation}
is finite. Then we may mimic the structure kernel through the
approximation
\begin{equation}
\tilde{S}({\bf r},{\bf r}',z;\dlE) = \bar{S}\, \dlE \,\delta ( {\bf
r} - {\bf r}' )   \label{local}
\end{equation}
for  $|{\bf r}-{\bf r}'|$ much greater than the range of $S({\bf
r}'')$. Let us take $\dlE$ small enough for the density matrix
element between states $m$ and $n$ with $(E_n-E_m)=\dlE$ to be, on
average, approximately independent of position and energy,
\begin{equation} \label{energy-independent}
| \Psi^*_n ( {\bf r} ) \Psi_m ( {\bf r} ) |^2 \simeq \frac{1}{A^2}
\ .
\end{equation}
This set of approximations leads to a transition rate from state
$n$ to $m$ with the particularly simple expression:
\begin{eqnarray}
\Gamma_{mn} &=& \bar{S}\,\dlE \int_A d {\bf r}\, | \Psi^*_n (
{\bf r} ) \Psi_m ( {\bf r} ) |^2 \label{local_1} \\
&\simeq& \frac{\bar{S}}{A} \dlE \ . \label{local2}
\end{eqnarray}

Within this regime of validity, the width of a quasiparticle state
of energy $E$ can be written as
\begin{equation}
\Gamma ( E ) = \frac{\bar{S}}{A} \nu(E)\int_{\Delta_0}^{E - E_F} d
( \dlE )\, \dlE \propto ( E - E_F )^2\ , \label{gamma_local}
\end{equation}
where the density of states $\nu(E)$ is assumed to be vary slowly
on the scale of $E-E_F$ (an exact property for a two-dimensional
dot).

In practice, $\bar{S}$ defined in (\ref{bar-S}) is never finite,
as can be seen by noting that $\bar{S}=\lim_{\dlE \rightarrow 0}
S({\bf q}=0, z, \dlE)/\dlE$ and deriving $S(q,z,\dlE)$ for any
reasonable screening model (see the Appendix for specific cases).
Fortunately, the requirements to obtain (\ref{local2}) and
(\ref{gamma_local}) are much less stringent than the finiteness of
(\ref{bar-S}). While the electronic behavior of a dot very close
to a metallic gate deserves a careful study, a preliminary
investigation reveals the following results: A semi-infinite
three-dimensional gate yields Fermi liquid behavior for $z\ll L$,
with logarithmic corrections if it is dirty. The behavior of a dot
close to a gate made of a two-dimensional electron gas is somewhat
more singular. We find marginal Fermi liquid behavior [$\Gamma(E)
\propto (E-E_F)$] if the 2D gate is diffusive, and Fermi liquid
behavior with logarithmic corrections if it is ballistic. We must
note that these calculations do not include additional screening
by the dot, which may be important for $L > \lambda_{\rm TF}$. For
such large dots, we expect to find full Fermi liquid behavior for
both two-dimensional and semi-infinite three-dimensional gates.

The result (\ref{gamma_local}) has the form which corresponds to a
{\it standard Fermi liquid} \cite{AFS82}. The effect of the
external environment is then indistinguishable from that of the
intrinsic charge fluctuations of the dot, which are not considered
in the present analysis but which should dominate in the limit of
very large dots. The estimation of the typical matrix element in
Eq. (\ref{local_1}) can be extended to diffusive dots\cite{M00},
leading to logarithmic corrections to Eq. (\ref{gamma_local}).
Similar effects for bulk diffusive systems are well known in the
literature\cite{LR85}. Those calculations \cite{M00,LR85} have
been made in the context of studies of the effect of intrinsic
charge fluctuations.

\section{Screened dipole.}
\label{sec:screeneddip}

For dots far away from the gate and large enough to partially
screen the external electric field ($z \gg L \gg \lambda_{\rm
TF}$) an intermediate regime should be considered. The potential
which induces the electronic transitions in the dot has now to be
taken as that of an external uniform electric field screened by
the two-dimensional electron gas of the dot. Since the dot is
two-dimensional, the screening length is independent of the
electron density. In addition, in contrast with the
three-dimensional case, the electric field penetrates beyond this
length. Near the boundary, the charge density and the screened
potential, $V_{\rm TF} ( {\bf r} )$, depend algebraically on the
distance to the edge \cite{A00,ABG02}, ${\xi}$, as ${\xi}^{-1/2}$.
This power-law behavior allows us to estimate the dependence on
the energy difference, $\dlE$, of the transition matrix elements,
to leading order in $\lambda_{\rm TF} / L$. For a ballistic
quantum dot, dimensional analysis applied to the $ {\xi}^{-1/2}$
scaling of the potential yields
\begin{equation}
\int_A d {\bf r} \,  \Psi_n^* ( {\bf r} ) \Psi_m ( {\bf r} )
V_{\rm TF} ( {\bf r} ) \propto
(E_n - E_m)^{-1/2} \ . \label{scr_dipole}
\end{equation}
Such a classical estimation of the matrix elements can be carried
out exactly for the case of a disc \cite{AW93} or a rectangular
strip \cite{SM01} unbounded in one direction.  As in those cases,
we expect the scaling in Eq. (\ref{scr_dipole}) to be valid also
for any ballistic chaotic quantum dot. This behavior of the matrix
element leads to a lifetime:
\begin{equation}
\Gamma ( E ) \propto E - E_F \ ,
\end{equation}
i.e., the lifetime of a quasiparticle is proportional to the
energy of the quasiparticle itself. This type of dependence is
usually associated to {\it marginal Fermi liquid behavior} in bulk
systems.

Using the results in the Appendix, a more quantitative estimate
yields
\begin{equation}
\Gamma ( E ) \approx \frac{\lambda_{\rm TF}^2}{z^d}
\frac{E-E_F}{(k_F')^{d-2} k_F l'} \label{gamma_scr} \ .
\end{equation}
Like in the unscreened dipole case, interference among many
trajectories leads to $\Gamma(E)\simeq 0$ for $(E-E_F)\ll \hbar
v_F/L$.

\section{Discussion.}
\label{sec:conclusion}

We have calculated the lifetime of electronic states in ballistic
quantum dots due to the presence of metallic gates. Depending on
the size of the dot ($L$), the dot-gate distance ($z$), and the
screening length within the dot ($\lambda_{\rm TF}$), we find
different regimes, schematically described in Table I.
\begin{table}
\begin{tabular}{||cc|c|c||}
\hline \hline
&$\lambda_{\rm TF}/L$ &$\gg 1$ &$\ll 1$ \\
$z/L$ & & &
\\ \hline
$\gg 1$  & &dipole, NFL (III) &screened dipole, MFL (V)
\\ \hline
$\ll 1$  & &short range, FL (IV) &short range, FL (IV)
\\  \hline  \hline
\end{tabular}
\caption{Sketch of the regimes studied in the text. FL stands for
Fermi liquid behavior, while NFL and MFL indicate non-Fermi liquid
and marginal Fermi liquid, respectively. The roman numbers
indicate the sections where the various cases are discussed in the
main text.}
\end{table}
The level width depends quadratically on the quasiparticle
energies when the screening is short ranged. This happens when the
dot is sufficiently close to the metallic gate or for large dots. Such
behavior implies the existence of well defined quasiparticles inside
the dot, consistently with the Fermi liquid theory.

We find deviations from this behavior when the dot is sufficiently
far from the gate, so that the charge fluctuations at the gate
induce an almost uniform electric field at the dot. This field may
be either unscreened ($L \ll \lambda_{\rm TF}$) or imperfectly
screened ($\lambda_{\rm TF} \ll L$), because of the
two-dimensional nature of the dot here considered. In the absence
of screening, the coupling between the electrons in the dot and
the external gates is correctly described by the standard
Caldeira-Leggett model of a particle coupled linearly in its
position to a bath of oscillators. The lifetime then shows a
plateau at energies $E-E_F$ smaller than $E_F$, but still larger than
$\hbar v_F/L$ (see Fig. (\ref{tau})). This corresponds to non--Fermi-liquid
behavior. When the dipole induced by the gate is (imperfectly) screened,
the level width is linear in the energy. Then the definition of the
quasiparticle peaks is not enhanced near the Fermi energy as
strongly as in the case of a Fermi liquid. This is the case of a
marginal Fermi liquid.

Remarkably, the effect of the gate is enhanced for gates of
reduced dimensions, as revealed by the $z^{-d}$ dependence of the
linewidths. In this sense, it is important to note that the
effective dimensionality $d$ of the response of the gate is
determined by the ratio of its various dimensions to the distance
to the dot. For instance, the potential fluctuations induced at a
distance $z$ by a metallic wire of diameter much smaller than $z$
can be described as if the wire were one-dimensional (see the
Appendix), even if the wire is fully three-dimensional, with its
diameter much larger than its Fermi wave length.

Inspection of Table I suggests that the degree of departure from
Fermi liquid (FL) behavior seems to go through a maximum as a
function of the dot-gate distance: At small distances, short-range
correlations dominate causing FL behavior to prevail, as shown in
section IV. This limit is essentially equivalent to that in which
the dot merges into the gate, with global FL behavior. At longer
distances, the analysis of sections III and V indicates a
departure from the FL regime, especially for small dots. At much
longer distances, one expects the interaction between dot and gate
to be negligible. Then internal, short-range correlations dominate
within the dot and FL properties are again recovered. We note that
this apparently reentrant behavior is compatible with the
monotonous decay of the electronic linewidth with distance [See.
Eq. (\ref{general-gamma-mn}], since the prefactor of the quadratic
energy dependence characteristic of FL behavior are very large in
the case of short distances.

It is interesting to compare the estimates obtained here with the
lifetimes expected from the decay of the quasiparticles into
internal excitations of the dot. That calculation, for a
disordered dot, gives\cite{SIA94}:
\begin{equation}
\Gamma ( E ) \simeq \frac{1}{k_F \lambda_{\rm
TF}}\frac{(E-E_F)^2}{E_F } \label{diffusive} \ .
\end{equation}
Comparing this expression with the contribution from the gate, Eq.
(\ref{gamma_scr}) (we assume $L \gg \lambda_{\rm TF}$, i.e., the
screened case), we find that the effect of the gate dominates if
\begin{equation}
E -E_F< \frac{\hbar v_F}{L} \left( \frac{L}{z} \right)^d \frac{(
k_F \lambda_{\rm TF} )^3 ( k_F' L )^{2-d}}{( k_F l' ) ( k_F L )}\
.
\end{equation}
This inequality is based on the result (\ref{gamma_scr}), which
only applies for $(E-E_F)> \hbar v_F/L$, being zero below that
energy. Thus, the dominance of gate effects over a significant
energy range requires:
\begin{equation}
(k'_F z)^d < \frac{L}{l'}\,\frac{(k_F \lambda_{\rm
TF})^3}{(k_F/k'_F)^2} \ .
\end{equation}
Since, in turn, the dipole approximation requires $z \gg L$, one
is led to the condition:
\begin{equation}
\begin{array}{lclr}
k'_F L &\ll &\left[\frac{(k'_F/k_F)^2}{k'_F l'} (k_F \lambda_{\rm
TF})^3 \right]^{1/(d-1)} \  &d > 1 \nonumber \\
1 &\ll &\frac{(k'_F/k_F)^2}{k'_F l'} (k_F \lambda_{\rm TF})^3
 \  &d = 1 \end{array}
\end{equation}
Since the quantity with square brackets is $\sim 10^2$, we
conclude that the observation of the effects discussed here is
entirely possible if the gate is effectively one-dimensional ($z
\ll W$, where $W$ is the width of the wire). The optimal range of
distance is, roughly, $L < z < 100 L$. On the other hand, the
observation would be difficult for a two-dimensional gate, since
it would require $k'_F L < 10^2$, and practically impossible for a
semi-infinite 3D gate.

The unscreened regime, $\lambda_{\rm TF} \gg L$, corresponds to
dots smaller than $\hbar^2 \epsilon_0  / m^* e^2 $, which, for
realistic values of the parameters, $\epsilon_0 \sim 12 ,  m^*
\sim 0.06\, m_e$, corresponds to $L < 10^2$\AA. In addition, the
calculations presented here require the number of electrons in the
dot to be much larger than one. These dots require electronic
densities of order $\sim 10^{14}$ cm$^{-2}$. This value is much
higher than those achieved with present day techniques.

For sufficiently large dots, the mean free path $l$ becomes smaller
than the size $L$, and we have a crossover from the ballistic regime
considered here to a diffusive one. In diffusive
systems, typical matrix elements involving states $| m \rangle$
and $| n \rangle$ do not depend on the energy difference $E_n -
E_m$ for energies such that $| E_n - E_m | \ll E_c$, where $E_c =
 \hbar D  / L^2$ is the Thouless energy. This corresponds to an effective
short-range interaction behavior, as discussed in Sec. IV [see Eq.
(\ref{energy-independent})]. Thus, in the disordered regime, the
anomalous effects discussed in this paper will be cut off at
energies of order $E_c$, i.e., we will have conventional Fermi
liquid lifetimes for $E-E_F < E_c$.

The small sizes and the two-dimensional character of the quantum
dot are essential for the non-Fermi liquid results that we have
obtained. A three dimensional dot with $\lambda_{\rm TF}<L$ would
result in a complete screening of the gate fluctuations, putting
us in the short range regime and thus yielding a Fermi-liquid type
of quasiparticle lifetime. For one-dimensional systems, the decay
rate of a single particle level of quantum number $n$, coupled to
a Caldeira-Leggett environment, has been shown\cite{IJR01} to scale
linearly with $n$. The naive extension of such a universal
behavior for fermionic systems would also lead to an electronic
lifetime that exhibits a plateau as a function of $E - E_F$.
However, our starting point of well defined quasiparticles would
not be appropriate in one dimension.

Finally, we wish to stress that, due to the reciprocal character
of the microscopic interactions, in those cases where have found a
departure from Fermi liquid behavior in the dot, a similar
deviation will be realized locally in the region of the metallic
gate which is most strongly affected by the presence of the
quantum dot.

Possible experiments to measure the electronic lifetimes here
studied may include transport spectroscopy of quantum dots
\cite{ihn03,sigr03} or, through their effect on the pattern of
standing waves, STM studies of the electron density within the dot
similar to those performed for quantum corrals
\cite{mano00,fiet03}.

\section{Acknowledgments.}
We are grateful to G.-L. Ingold and F. von Oppen for useful
discussions. This work has been supported by the EU TMR Program
under Contract No. HPRN-CT-2000-00144. The support of the
Ministerio de Ciencia y Tecnolog\'{\i}a (Spain) under Grants No.
BFM2001-0172 and MAT2002-0495-C02-01 is also acknowledged.

\appendix

\section{Dipolar coupling to a metallic gate.}
In this appendix we estimate the coefficient ${S}_0$ defining the
structure factor in the dipole approximation. For specificity, we
will assimilate the metallic gate to a disordered electron layer.
The function $S ( {\bf {r}} , {\bf {r}}' ; E )$ defined in Eq.
(\ref{structure}), where ${\bf {r}}$ and ${\bf {r}}'$ are
positions within the dot, can be written, in general, as
\begin{equation} S({\bf {r}},{\bf {r}}',z;E) = -{\rm
Im} \, V_{\rm scr} ({\bf {r}} - {\bf {r}}',z;E/\hbar )
\end{equation}
where $V_{\rm scr}$ is the screened interaction induced by the
gate. For a two dimensional gate located at a distance $z$ from
the dot, the Fourier transform of $V_{\rm scr}$ is:
\begin{equation}
V_{\rm scr} ( {\bf {q}} ,z; \omega ) = \frac{2 \pi e^2\, e^{- 2q
z}}{q\, \epsilon ( {\bf {q}} , \omega )} \ ,
\end{equation}

\nin where ${\bf {q}}$ is a two-dimensional vector in the plane of the
dot. For a diffusive 2D electron gas, we have:
\begin{equation}
\epsilon({\bf {q}} , \omega ) = 1 + \frac{2 \pi e^2}{q}\frac{D
\nu_2 q^2}{D q^2 - i \omega}
\end{equation}
where $D = \hbar^2 k_F' / m l'$ is the diffusion coefficient, $l'$
the mean free path, and $\nu_2 = m / \pi\hbar^2$ is the
density of states of the two-dimensional gate. In the long
wavelength limit, we obtain
\begin{equation}
{\rm Im} \, V_{\rm scr} ( {\bf {q}} ,z ; \omega ) \simeq -
\frac{e^{- 2 qz}   \,\omega }{D \nu_2 q^2} \ ,
\end{equation}
where we have assumed $\omega \ll Dq^2$. The exponential $\exp(- 2
qz )$ implies that only wave-vectors such that $q\ll z^{-1}$
contribute to $V_{\rm scr}({\bf {r}} - {\bf {r}}',z;\omega)$.
Assuming that the typical distances within the dot are such that
$|{\bf {r}} - {\bf {r}}' | \ll z$, we can write:
\begin{equation}
{\rm Im}\, V_{\rm scr} ( {\bf {r}} - {\bf {r}}' , z;\omega )
\simeq - \int \frac{d^2 q}{8\pi^2}  \frac { | {\bf {q}}\cdot (
{\bf {r}} - {\bf {r}}' ) |^2 e^{- 2 qz } \, \omega } {D \nu_2 q^2}
\label{v_scr_2D}
\end{equation}
which leads to the dipolar approximation discussed in the
text \cite{comm1}.

The calculation of ${\rm Im}\, V_{\rm scr}({\bf {r}} - {\bf
{r}}',z;\omega )$ for a semi-infinite {\it three-dimensional} gate
can be carried out in a similar way if we assume that
quasiparticles within the gate are specularly reflected at the
boundary \cite{RM66}. This approximation has been widely used in
the literature for the study of the related problems of energy
dissipation by moving charges \cite{sols88} or the decay of image
states at metallic surfaces\cite{EFS85,Eetal00}. Then, the
effective dielectric function which describes the effect of
electrostatic screening by the gate at points outside the gate can
be written as a two-dimensional integral over the surface of the
gate. This approximation becomes exact at large distances. The
resulting screened potential outside the gate is:
\begin{equation}
V_{\rm scr}({\bf q},z; \omega ) = e^{- 2 qz} \frac{2 \pi e^2}{q}
\frac{B ( {\bf {q}} , \omega)-1}{B ( {\bf {q}} , \omega )+1} \ ,
\label{v_scr_3D}
\end{equation}
where
\begin{equation}
B ( {\bf {q}} , \omega ) \equiv \frac{q}{\pi} \int \frac{d
q_z}{(q^2+q_z^2)\epsilon ( {\bf {q}} , q_z ; \omega )} \ ,
\end{equation}
$\epsilon ( {\bf {q}} , q_z , \omega )$ being the bulk dielectric
function. Using, as for the two-dimensional gate, the dielectric
function of a diffusive electron liquid, and expanding for $| {\bf
{r}} - {\bf {r}}' | \ll z$, we find
a result identical to Eq. (\ref{v_scr_2D}) with $\nu_2 q^2$ in the
denominator of the integrand replaced by $\nu_3 q$, where $\nu_{3}
=m k_F'  / \pi^2\hbar^2$ is the density of states of the
three-dimensional gate.

Finally, it is straightforward to generalize the previous
calculations to the case where the gate is a one-dimensional
metallic gate. The main difference is that the integral over the
positions at the gate is one-dimensional.

Using eqs.(\ref{v_scr_2D}) and (\ref{v_scr_3D}) we can estimate
the value of ${S}_0$ in Eq. (\ref{matrix_1}). We find, apart from
factors of order unity,
\begin{equation}
{S}_0^{-1}(z) \approx \left\{ \begin{array}{lr}
z l' \, \, \, \, \, &1{\rm D} \\
z^2 ( k_F' l' ) \, \, \, \, \, &2{\rm D}\\
z^2 ( k_F' z ) ( k_F' l' ) \, \, \, \, \,  &3{\rm D}
\end{array} \right. \label{prefactor}
\end{equation}
Corrections of order $\log ( z/ R )$, where $R$ is the radius of
the gate, are neglected when the gate is effectively
one-dimensional.


\begin{references}
\bibitem{comm1}
The factor $|{\bf q}\cdot({\bf r} - {\bf r}')|^2$ in the integrand
is equivalent to $q^2(r^2+r'^2-2{\bf r}\cdot{\bf r}')/2$. Since
the term $(r^2+r'^2)$ does not contribute to the decay rate in Eq.
(\ref{lifetimes}), we obtain effectively the bilinear behavior
$S({\bf r},{\bf r}')\sim {\bf r}\cdot{\bf r}'$ which characterizes
the coupling for $z \gg L$.
\end{references}
\end{document}